\documentclass[preprint]{aastex}
\def\sss{\scriptscriptstyle}
\def\^#1{^{\sss #1}}
\def\_#1{_{\sss #1}}
\def\beq{\begin{equation}}
\def\eeqno#1{\label{#1}\end{equation}}
\def\ten#1#2{^{\sss#1}_{\sss#2}}

\def\az{a\_{0}}

\def\l0{\ell\_{0}}

\def\rar{\rightarrow}

\def\l{\lambda}

\def\rp{\rho_p}

\def\f{\phi}
\def\fN{\phi\^N}

\def\fs{\f^*}
\def\gfs{\grad\fs}

\def\k{\kappa}
\def\z{\zeta}

\def\r{\rho}
\def\rh{\hat\rho}
\def\m{\mu}
\def\n{\nu}
\def\Up{\Upsilon}
\def\C{\Gamma}

\def\L{\mathcal{L}}

\def\Q{\mathcal{Q}}

\def\M{\mathcal{M}}

\def\Mt{\tilde\M}

\def\D{\Delta}

\def\d{\delta}
\def\drt{d^3r}
\def\a{\alpha}
\def\b{\beta}
\def\c{\gamma}
\def\d{\delta}

\def\vr{{\bf r}}

\def\vR{{\bf R}}

\def\vF{{\bf F}}

\def\vD{{\bf D}}
\def\vv{{\bf v}}

\def\va{{\bf a}}

\def\vf{{\bf f}}
\def\vF{{\bf F}}

\def\vP{{\bf P}}

\def\S{\Sigma}

\def\Sz{\Sigma_0}
\def\grad{\vec\nabla}
\def\div{\vec \nabla\cdot}
\def\gf{\grad\phi}
\def\fpg{4\pi G}

\def\gmn{g\_{\m\n}}
\def\Gmn{g\^{\mu \nu}}

\def\hgmn{\hat g\_{\m\n}}
\def\hgh{\hat g^{1/2}}
\def\gh{g^{1/2}}

\def\gft{\grad\tilde\f}
\def\fh{\hat\f}
\def\gfh{\grad\fh}

\def\ft{\tilde\f}
\def\fb{\bar\f}
\def\gfb{\grad\fb}

\def\emn{\eta\_{\m\n}}
\def\rph{\hat\rp}
\def\mi{m\_i}
\def\mhk{\hat m\_k}
\def\E{E(\vr\_1,...,\vr\_{N})}
\def\El#1{E(#1\vr\_1,...,#1\vr\_{N})}

\begin{document}

\title{Matter and twin matter in
 bimetric MOND}
\author{Mordehai Milgrom }
\affil{ The Weizmann Institute Center for Astrophysics}
\begin{abstract}
Bimetric MOND (BIMOND) theories, propounded recently, predict
peculiar gravitational interactions between matter and twin matter
(TM). Twin matter is the hypothetical matter that might couple
directly only to the second metric of the theory, as standard matter
couples only to the first. Considerations of cosmology in the BIMOND
framework suggest that such TM might exist and copy matter in its
attributes. Here I investigate the indirect interactions that BIMOND
theories predict between local, nonrelativistic mass concentrations
of matter and TM. The most salient result is that in the deep-MOND
regime of the matter-TM-symmetric theories, TM behaves as if it has
a negative gravitational mass relative to matter (active and
passive, with the inertial mass still positive). To wit, interaction
within each sector is attractive MOND gravity, but between matter
and TM it is repulsive MOND gravity. Using the space-conformal
invariance of the theory in the deep-MOND regime, I derive various
exact results; e.g., the repulsive force between a matter and TM
point masses (space-conformal theories are a natural framework for
masses of opposite signs). In the high-acceleration regime, the
interaction depends on a parameter, $\b$ (the strength of the
Einstein-Hilbert action for matter). For the favored value $\b=1$,
matter and TM do not interact in this regime; for $\b<1$ they
attract; and for $\b>1$ they repel each other. Such interactions may
have substantial ramifications for all aspects of structure
formation, such as matter distribution, peculiar velocities, and
effects on the CMB. The repulsive interactions probably lead to
segregation of matter and TM structures, leading, in turn, to
intermeshing of the respective cosmic webs, with high-density nodes
of one sector residing in the voids of the other (possibly conducing
to efficient evacuation of the voids). Weak gravitational lensing by
TM seems the best way to detect it or constrain its attributes. In
the MOND regime a TM body acts on matter photons as a diverging
lens. Strong lensing occurs in the high acceleration regime, and
thus depends on $\b$. For $\b=1$, a TM mass does not bend (matter)
light in the high-acceleration regime: no strong lensing effects of
TM are expected in this case. I also discuss briefly asymmetric
theories.
\end{abstract}

\maketitle
\section{Introduction}
Bimetric theories of gravity involve two metrics as independent
degrees of freedom: one felt directly by standard matter, $\gmn$,
and an auxiliary metric, $\hgmn$. Aspects of such theories have been
extensively discusses; e.g., by Isham, Salam, \& Strathdee (1971),
by Rosen (1974), and, for some recent treatments, with references to
other and to earlier work, see, e.g., Boulanger \& al. (2001),
Damour \& Kogan (2002), Blas, Deffayet, \& Garriga (2006),
Ba\~{n}ados, Ferreira, \& Skordis (2009), and Ba\~{n}ados, \& al.
(2009). It may be the case that the metric $\hgmn$ is indeed merely
an auxiliary field in the description of gravity of standard matter.
It is, however, natural to suppose that $\hgmn$ comes with a matter
sector of its own, and should be viewed on a par with $\gmn$. This
opens the way for introducing twin matter (TM), which may couple to
its ilk in the same way as standard matter does
(electromagnetically, weakly, etc.), but which couples
gravitationally only to $\hgmn$, just as matter couples only to
$\gmn$. Since the two metrics are coupled through a gravitational
term in the action, matter and TM do couple indirectly by some
unconventional gravitational interaction.
\par
A bimetric formulation of MOND (BIMOND) has been propounded recently
(Milgrom 2009b), which points even more forcibly to the possible
existence of TM.
\par
BIMOND is governed by an action of the form
 \beq I=-{1\over 16\pi G}\int[\b \gh R +\a\hgh \hat R
 -2(g\hat g)^{1/4}f(\k)\az^2\M(\bar\Up/\az^2)]d^4x
 +I\_M(\gmn,\psi_i)+\hat I\_M(\hat g\_{\m\n},\chi_i).  \eeqno{gedat}
Here $\bar\Up$ represents a collection of scalar variables formed by
contractions of the acceleration-like tensors
$C\^\a\_{\b\c}=\C\^\a\_{\b\c}-\hat\C\^\a\_{\b\c}$, where
$\C\^\a\_{\b\c}$ and $\hat\C\^\a\_{\b\c}$ are the Levi-Civita
connections of the two metrics. Also, $\k\equiv(g/\hat g)^{1/4}$
($g$ and $\hat g$ are minus the determinants of the two metrics),
$G$ is the phenomenological Newton constant, and I use unites where
$c=1$. ($\M$ may also depend on scalars constructed from the two
metrics, such as $\k$.) $I\_M$ is the action for matter, whose
degrees of freedom and their derivatives are collectively marked by
$\psi_i$; these interact among themselves, and couple to the metric
$\gmn$ alone in the standard way, and similarly $\hat I\_M$ is the
TM action. I have worked, in particular, with favorite choices of a
the scalar variable in $\M$ built from the tensor
 \beq\Up\_{\m\n}=C\ten{\c}{\m\l}C\ten{\l}{\n\c}
-C\ten{\c}{\m\n}C\ten{\l}{\l\c}; \eeqno{mulpat} for example,
 \beq \bar\Up=-{1\over 2}
\Gmn\Up\_{\m\n}.\eeqno{scalala}
\par
To obviate possible confusion, especially in the present MOND
context, I note at the outset that the TM is not the putative dark
matter (DM), and is not taken to play its role of enhancing gravity
in galactic systems. It is still MOND departure from standard
general relativity (GR) that replaces dark matter. TM may still
produce some effects, e.g., in structure formation, that are
conventionally attributed to cosmological DM. It may also linger in
otherwise matter-dominated territory to produce some visible effects
(see \ref{other}).
\par
We have no a priori idea of how the TM sector looks like; neither do
we have any observational information on the subject. Does it exist
at all? Is it made of the same `stuff' as matter, and is it subject
to the same physics? Is it present in the Universe in the same
amounts? Has it undergone similar processes in cosmic history (big
bang, inflation, seed fluctuations, structure formation, etc.)? We
do not even know whether matter coupling to gravity is the same in
the two sectors. To be able to progress I shall assume symmetry
between the sectors to the effect that TM duplicates matter in its
composition, interactions, etc., and that if $(\hgmn,\chi_i)$ is the
same configuration as $(\gmn,\psi_i)$, then
$I\_M(\gmn,\psi_i)=(\a/\b)\hat I\_M(\hat g\_{\m\n},\chi_i)$. This
ensures that in the absence of the interaction $\M$, the dynamics
within the two sectors are identical. With the interaction the above
assumptions do not ensure complete symmetry when $\a\not =\b$.
\par
In cosmology, the above assumptions plus the assumptions that the
cosmic matter contents of the two sectors are the same, and a
symmetric choice of $f(\k)=(\b\k+\a\k^{-1})/(\a+\b)$, was shown
(Milgrom 2009b) to lead to cosmologies in which $\hgmn=\gmn$, with
each metric describing a standard, Friedmann-Robertson-Walker (FRW)
cosmology, with a cosmological constant
$\Lambda=-\az^2\M(0)/(\a+\b)$. This automatically retains the known
successes of such a cosmology. Cosmology is thus, arguably, the
strongest motivation for postulating that TM exists. In such BIMOND
cosmologies, $G/\b$ appears in the matter sector of such a cosmology
as Newton's constant. This may induce us to prefer $\b\approx 1$.
The value $\b=1$ is also special because such BIMOND theories have a
simple limit  when $\az\rar 0$: they tend to GR in the matter sector
when $\M'(z)\rar 0$ for $z\rar\infty$, which is required for the
nonrelativistic (NR) limit to yield Newtonian dynamics for $\az\rar
0$. I have not yet investigated the limit $\az\rar 0$ for BIMOND
theories with $\b\not=1$ to see to what extent they differ from GR.
I will, none the less, keep the discussion more general, taking
$\b=1$, as an example, only at a later point.
\par
While on cosmological average the two metrics might be the same,
departures from equality must occur due to random fluctuations in
density, which are different in the two sectors. To treat
large-scale-structure formation through the development of small
perturbations on the background of the expanding Universe (including
their imprints on the CMB, etc.) we need to expand the BIMOND
equations of motion around the double FRW metric describing
cosmology at large.
\par
Note, importantly in this context, that even if early on the
Universe is characterized by high accelerations (as measured, e.g.,
by $cH$, where $H$ is the Hubble parameter), what determines whether
we are in the MOND regime is the argument of the interaction
function $\M$. This argument is small when the two metrics are near
each other. Thus, structure formation in the initial stages, in an
otherwise symmetric universe, occurs in the MOND regime. So, even at
small amplitude of fluctuations the development is nonlinear, as the
MOND potentials are not linear in the density fluctuations, right
from the outset. Structure formation, in all its aspects, such as
the distribution of matter, peculiar motions, and CMB fluctuations,
are thus phenomena that might be greatly affected by the new physics
inherent in BIMOND, in particular by the possible effects of TM. The
treatment of this problem has to be left to numerical simulations.
\par
In this paper I take up the more modest task of investigating the
dynamics of well formed systems involving  well separated masses of
matter and TM, of sizes much smaller than cosmic scales. These are
then taken to be NR systems on a double Minkowski background
($\hgmn\approx\gmn\approx\emn$). The appropriate limit of BIMOND
theories under these circumstances was shown to be the class of NR
MOND theories described in detail in Milgrom (2009c).
\par
The analysis here includes derivation of the forces between
combinations of matter and TM bodies, and also the characteristics
of gravitational lensing of matter photons by TM.
\par
In section \ref{formalism}, I describe the field equations that
govern NR interactions within and between the two matter sectors. In
section \ref{symmetric}, I specialize to the class of theories that
are fully symmetric in matter and TM. In section
\ref{observational}, I consider some possible observational
consequences of the existence of TM and its peculiar interaction
with matter, including gravitational lensing. In section \ref{asy},
I look briefly at examples of asymmetric theories. Section
\ref{disc} is a discussion.

\section{\label{formalism}Formalism}
For a system of slowly moving matter and TM distributions  $\r$ and
$\rh$, respectively, on a double Minkowski background, for the
choice of variable $\bar\Up$ as in eq.(\ref{scalala}), the solution
for the metrics of the BIMOND field equations is (in some gauge)
  \beq \gmn=\emn-2\f\d\_{\m\n},~~~~
 \hat g\_{\m\n}=\emn-2\fh\d\_{\m\n}
\eeqno{rukun} (Milgrom 2009b).\footnote{Choices of the acceleration
scalar arguments of $\M$ other than that given in eq.(\ref{scalala})
result in other NR limits of the corresponding BIMOND theory than
those given by eq.(\ref{rukun}). In general, more potentials are
needed to describe each metric, and a more complicated set of
coupled equations for these have to solved (Milgrom 2009b).} Here,
the potentials $\f$ and $\fh$ are solutions of the NR limit of the
BIMOND equations; these equations can be derived from the Lagrangian
$L=\int\L\drt$, with
   \beq \L=-{1\over 8\pi
G}\{\a(\gfh)^2+\b(\gf)^2-\az^2\M[(\gf-\gfh)^2/\az^2]\}
  +\r({1\over 2}\vv^2-\f)+
  \c\rh({1\over 2}\hat\vv^2-\fh).  \eeqno{futcol}
The last two terms in the Lagrangian density tell us that $\f$ is
the (MOND) gravitational potential for matter, and $\fh$ is that for
TM, in the sense that the acceleration of matter and TM test
particles is given by $\dot\vv=-\gf$, and $\dot{\hat\vv}=-\gfh$,
respectively.\footnote{This NR theory is, in fact, the NR limit of a
group of variations on the relativistic theory, where we can use in
the contraction in $\bar\Up$, $\hat g\^{\m\n}$ instead of $\Gmn$, or
use combinations of such variables, for more symmetry between the
two metrics.}

\par
Invariance of the theory to translations implies that for a closed
system we have a vanishing total force:
 \beq \vF=-\int(\r\gf+\c\rh\gfh)\drt=0,  \eeqno{kipola}
and the conserved momentum is
  \beq \vP=\int(\r\vv+\c\rh\hat\vv)\drt.  \eeqno{iopala}
\par
I define $\rh$ as nonnegative, so the sign of $\c$ matters. In the
spirit of what I said above about the relativistic theory, I take
from now on $\c=\a/\b$.\footnote{In the present context, we can
absorb $|\a/\b|$ in $\rh$; so, this choice of $|\c|$ may be viewed
as the choice of a convenient normalization for $\rh$.}
\par
The field equations are
 $$\D\f=\fpg\b^{-1}\r+\b^{-1}\div(\M'\gf^*)\equiv \fpg(\r+\rp),$$
  \beq \D\fh=\fpg\b^{-1}\rh-\a^{-1}\div(\M'\gf^*)\equiv
  \fpg(\rh+\rph),\eeqno{hugtal}
where $\fs=\f-\fh$, and $\rp$ and $\rph$ playing the role of
`phantom matter' (PM) densities for the two sectors. ($\a=0$, or
$\b=0$ do not give MOND theories, and I exclude such values.)
 Note that for a given configuration the amounts of
PM felt by matter and TM are, in general, different.
\par
Subtracting one equation from the other we get a decoupled equation
for $\fs$
 \beq\div[\m^*(|\gfs|/\az)\gfs]=\fpg(\r-\rh),  \eeqno{cutes}
 where
  \beq \m^*(x)\equiv\b-{\a+\b\over\a}\M'(x^2). \eeqno{cueto}
\par
There is an equivalent, but more transparent, way to write the
theory for the case $\a+\b\not=0$: Define
 \beq \Mt(z)=-\M(z/q)+{\a\b z\over (\a+\b)q},  \eeqno{gaduta}
where $q=\a^2/(\a+\b)^2$. Then, the Lagrangian density can be
written in terms of the potentials
 \beq \ft=\b\f+\a\fh,~~~~~\fb=\a\z(\f-\fh),   \eeqno{jubna}
where $\z\equiv (\a+\b)^{-1}$, as
   \beq \L=-{1\over 8\pi
G}\{\z(\gft)^2+\az^2\Mt[(\gfb)^2/\az^2]\}
  -\z\ft(\r+\a\b^{-1}\rh)-\fb(\r-\rh)
+{1\over 2}(\r\vv^2
  +{\a\over\b}\rh\hat\vv^2).  \eeqno{futmen}
The two potentials are now decoupled, satisfying the equations
 \beq \D\ft=\fpg(\r+{\a\over\b}\rh),~~~~~
 \div\{\Mt'[(\gfb/\az)^2]\gfb\}=\fpg(\r-\rh).  \eeqno{katpoy}
The potential $\ft$ is thus a linear combination
 \beq \ft=\fN+{\a\over\b}\hat\f\^N,  \eeqno{moreca}
where the Newtonian potentials in the two sectors, $\fN$ and
$\hat\f\^{N}$, are the solutions of the Poisson equation for
$\r$ and $\rh$ separately.
The matter and TM MOND potentials are then gotten as linear
 combinations
 \beq
 \f=\z\ft+\fb=(1+\l)\^{-1}(\a^{-1}\fN+
 \b^{-1}\hat\f\^N)+\fb,
 ~~~~~~~\fh=\z\ft-{\b\over\a}\fb=(1+\l)\^{-1}(\a^{-1}\fN+
 \b^{-1}\hat\f\^N)
 -\l\fb,
  \eeqno{rabutla}
where $\l=\b/\a$.
\par
The deep-MOND limit is formally implemented by taking
$\az\rar\infty$, $G\rar 0$,  with $G\az$  kept fixed. In this limit,
$\fb$ becomes super-dominant over the Newtonian-like potential
$\ft$. We then have to have $\Mt'(w)\rar w\^{1/2}$ giving for the
matter potential
 \beq \div(|\gf|\gf)=\fpg\az(\r-\rh),  \eeqno{katmer}
and $\fh=-(\b/\a)\f$ for the TM potential, so we can also write
  \beq \div(|\gfh|\gfh)=\fpg\az\left({\b\over\a}\right)^2(\rh-\r).
   \eeqno{katmpet}
\par
If we put matter and TM test particles at the same position, $\vr$,
in a gravitational field, the former will have an acceleration
$\va=-\gf(\vr)$, while the latter will be accelerated by
$-(\b/\a)\va$. (Matter and TM do not follow the same weak
equivalence principle, as is clear from the very construction of
BIMOND, with matter following geodesics of $\gmn$ and TM those of
$\hgmn$.) In the deep-MOND limit of theories with $\a$ and $\b$ of
the same sign, matter attracts matter as in standard MOND; TM
attracts TM as in MOND, but with an effective value of $G\az\rar
G\az(\b/\a)^2$; however, matter and TM repel each other.

\par
To get the Newtonian behavior in the matter sector for $\az\rar 0$;
namely, to get $\f\rar\fN$ in this limit, for a pure-matter
system,\footnote{Note that the Newtonian limit is defined by the
requirement that it reproduces Newtonian dynamics in the Matter
sector, not necessarily in the TM sector.} we have to
have\footnote{This has to be required only because we insisted that
$G$ is Newton's constant. Had we started from some general coupling,
$G'$, this relation would just constitute a definition of $G$ in
terms of $G'$.} $\Mt'(\infty)=1/(1-\z)$.
\par
Since $\Mt'(w)\approx w\^{1/2}$ for small $w>0$, and is thus
necessarily positive there, we cannot have $\z>1$, lest $\Mt'(w)$
vanishes at a finite value $w$, which it must not. This is the
condition derived in Milgrom (2009c) on the same grounds. We can
have $\z=1$ (and then $\Mt'\rar\infty$ at infinity); but this is
excluded by solar-system constraints. There may be other constraints
on the values of $\a,~\b$ from different consistency requirements in
the relativistic and NR theories, but such are yet to be found.
\par
With the above value of $\Mt'(\infty)$, we have in the
high-acceleration regime
 \beq \f=\fN+(\b\^{-1}-1)\hat\f\^{N},
 ~~~~~~\fh=\left[1+{(\b-1)(\b-\a)\over\a\b}\right]\hat\f\^{N}
 +{1-\b\over
 \a}\fN.  \eeqno{gumtas}
\par
We see that while $G$ is the gravitational constant in the Newtonian
limit of the matter sector, it is $\hat G=G[1+(\b-1)(\b-\a)/\a\b]$
that plays this role in the TM sector. For $\b=1$ or $\b=\a$ we have
$\hat G=G$ but otherwise they are different. In fact, for some
choices of $\a,~\b$ we have $\hat G<0$, in which case the $\az\rar
0$ limit corresponds to repulsive gravity in the TM sector. If we
deem this undesirable we can eliminate the corresponding $\a,~\b$
values. (For $\a,\b>0$ we have $\hat G>0$ with our already assumed
inequality $\z<1$.)
\par
We also see from eq.(\ref{gumtas}) that for $\b=1$, $\f=\fN$ and
$\fh=\hat\f\^{N}$, which means that there is no interaction between
the two sectors in the Newtonian regime. This is true for the fully
relativistic theory, where for $\b=1$ (irrespective of $\a$) BIMOND
separates in the limit $\az\rar 0$ to two copies of GR in the two
sectors [possibly with a cosmological constant $\sim
\az^2\M(\infty)$].
\par
When $\a\not=\b$ the dynamics in the two sectors can be quite
different. Such theories are worth investigating, but my purpose
here is not to conduct an exhaustive study of this class of
theories, only to demonstrate some salient results. To this end I
shall consider mainly theories that are fully symmetric in
matter-TM, namely those with $\a=\b$. I will then discuss briefly
some asymmetric cases.

\subsection{A quasi-linear formulation}
Solution of the field equations (\ref{katpoy}) requires solving a
nonlinear Poisson equation for the given matter-TM configuration at
hand. This may be rather taxing, especially when applying the theory
to time-dependent problems, such as that of large-scale-structure
formation. As in the case of the quasi-linear formulation of MOND
(QUMOND; Milgrom 2009c), which parallels the
Modified-Poisson-equation formulation of Bekenstein \& Milgrom
(1984), I describe here a quasi-linear theory, derivable from an
action, not equivalent to our theory, but which captures much of its
essence, and which should be much easier to apply. As in the case of
QUMOND, this requires adding an auxiliary potential $\psi$ to the
MOND potentials for matter and TM.\footnote{Since $\fb$ and $\ft$
are decoupled in the Lagrangian density in eq.(\ref{futmen}), we
simply apply to $\fb$ the same procedure that has lead to QUMOND.}
Consider the Lagrangian density
   \beq \L=-{1\over 8\pi
G}\{2\a\z(\gf-\gfh)\cdot\grad\psi+\z(\b\gf+\a\gfh)^2
-\az^2\Q[(\grad\psi/\az)^2]\} +\r({1\over 2}\vv^2-\f)+{\a\over
\b}\rh({1\over 2}\hat\vv^2-\fh).
   \eeqno{iii}
In terms of the two potentials $\fb$ and $\ft$ related to the MOND
potentials by eq.(\ref{jubna}) one can write
 \beq \L=-{1\over 8\pi
G}\{2\gfb\cdot\grad\psi+\z(\gft)^2 -\az^2\Q[(\grad\psi/\az)^2]\}
-\fb (\r-\rh)-\ft\z(\r+{\a\over\b}\rh)
  +{1\over 2}\r\vv^2+{\a\over 2\b}\rh\hat\vv^2,
   \eeqno{iiivv}
giving the field equations
 \beq \D\ft=\fpg(\r+{\a\over \b}\rh),~~~~~~\D\psi=\fpg(\r-\rh),
 ~~~~~~
  \D\fb=\div\{Q'[(\grad\psi/\az)^2]\grad\psi\}.  \eeqno{katreq}
It is easy to see that with the appropriate choice of $\Q$ [namely,
with $\tilde\M'(x^2)x=y$ being equivalent to $\Q'(y^2)y=x$] the
theory is a very good mimic of equations (\ref{katpoy}). For
example, in a spherically symmetric case they are identical.

\section{\label{symmetric}Dynamics in the fully symmetric theory}
In the fully matter-TM symmetric case, $\a=\b$, which seems to give
appealing cosmological solutions, we can write the field equations
(\ref{katpoy}) as
 \beq \D\ft=\fpg(\r+\rh),~~~~~
 \div\{\Mt'[(\gfb/\az)^2]\gfb\}=\fpg(\r-\rh),  \eeqno{katpet}
 and the MOND potentials are then given by
 \beq
 \f=\z\ft+\fb,~~~~~~~\fh=\z\ft-\fb,
  \eeqno{rabutret}
where now $\z=(2\b)^{-1}$.
\par
In the deep-MOND limit $\f$ satisfies eq.(\ref{katmer}), and
$\fh=-\f$.
\par
Interestingly, in the deep-MOND limit of the theory, TM behaves as
if it has a negative active and passive gravitational mass relative
to matter, while its inertial mass is still positive: Negative
active mass because $\rh$ enters the source for the matter $\f$
potential with a negative sign, and negative passive mass, because
it is accelerated by the gravitational field $\fh=-\f$. So bodies in
the same sector attract each other, while bodies in different
sectors repel each other. I discuss gravity in this important MOND
limit in more detail in subsection \ref{mond}.
\par
The case $\r\approx \rh$ has to be commented on. In this case
$\fb\approx 0$, while $\ft\approx 2\fN\approx 2\hat\f\^{N}$. The
MOND limit applies when $[G\az(\r-\rh)R]^{1/2}\gg G(\r+\rh)R$, where
$R$ is the characteristic size of the system. The configuration of
density near-equality is unstable, and with a small separation,
repulsion occurs (see subsection \ref{well}).\footnote{A similar
result applies in the more general, relativistic case: In
configurations where the energy-momentum tensors in the two sectors
are equal (with our normalization $\c=\a/\b$) the solution of the
field equations is $\hgmn=\gmn$, with both being the solution of the
standard Einstein equation for the configuration [with a
cosmological constant $\propto \az^2\M(0)$]. For example, a double
Schwarzschild metric is a spherically symmetric, vacuum solution of
the BIMOND equations. Matching it to an interior solution will show
that it corresponds to a central mass made of equal amounts of
matter and TM.}
\par
Consider now the Newtonian limit, $\az\rar 0$, where from the
general eq.(\ref{gumtas}) we have
 \beq \f=\fN+(2\z-1)\hat\f\^{N},~~~~~~\fh=\hat\f\^{N}+(2\z-1)\fN.
 \eeqno{jules}

\par
We see that, unlike the MOND limit, here the predicted fields do
depend on $\z$. Recall that for the choice, $\z=1/2~ (\b=1)$, we
have $\f=\fN$, $\fh=\hat\f\^{N}$, with each sector seeing exactly
its own Newtonian potential; so, matter and TM do not interact at
all. For $\z>1/2$ matter sees its own Newtonian potential plus a
fraction $2\z-1>0$ of that of TM; so they attract each other with a
reduced effective gravitational constant. For $\z<1/2$ they repel
each other.
\par
The case $\z\rar 0$ ($\b\rar\infty$) is also interesting: We see
from eq.(\ref{rabutret}) that in this case $\f=-\fh=\fb$ for the
full range of the theory (Newtonian-MOND); $\ft$ becomes immaterial,
and the MOND potential $\f$ satisfies the second of
eq.(\ref{katpet}). We saw that for $\z=0$ we have $\Mt'(\infty)=1$.
So in this case we end up with the MOND theory of Bekenstein \&
Milgrom (1984) with $\m(x)=\Mt'(x^2)$, but with TM entering the
theory as having a negative gravitational mass.
\subsection{Forces on bodies}
The force $\vF$ ($\hat\vF$) on a matter (TM) body that constitutes a
subsystem of the density $\r$ ($\rh$) within the volume $\upsilon$
($\hat \upsilon$) is
 \beq \vF=-\int\_{\upsilon}\r\gf\drt,
 ~~~~~~\hat\vF=-\int\_{\hat\upsilon}\rh\gfh\drt. \eeqno{kupala}
Because the theory is nonlinear we cannot use in these expressions
the potential produced by the system with the body in question
excluded (as can be done in the linear case). As a result, even for
a point mass $m$ at position $\vr$ we cannot write the force simply
as $\vF=-m\gf(\vr)$, where $\f$ is produced by the rest of the
system. In fact, the force is not even linear in the mass of the
body, and becomes so only for test particles. In Milgrom (1997,
2002a), I discussed in detail general properties of forces in such
nonlinear theories. Here I essentially use the results from these
papers.
\par
On dimensional grounds we can write the force between two matter
point masses, $M$ and $m$, a distance $r$ apart, as $F=-\az
Mf(m/M,r/R\_M)$, where $R\_M=(MG/\az)^{1/2}$ is the MOND radius for
the mass $M$, and a negative sign signifies attraction. The same
expression applies to the force between two TM point masses. For a
matter point mass $M$ and a TM point mass $\hat M$ we can write
$F=\az M f\^*(\hat M/M,r/R\_M)$.
\par
In the test-particle limit, where one of the masses is much smaller
than the other, $f(q,\l)$ and $f\^*(q,\l)$ are easy to obtain for
all values of $\l$, because the force on a test particle equals its
mass times the gradient of the potential produced by the massive
ponit mass, which can be gotten analytically. In the Newtonian limit
$\l\ll 1$ we see form eq.(\ref{jules}) that $f(q,\l\ll 1)\approx
q\l^{-2}$, which reproduces the Newtonian force, and $f\^*(q,\l\ll
1)\approx (2\z-1)q\l^{-2}$. Expressions for these functions in the
deep-MOND limit $\l\gg 1$ will be given in subsection \ref{mond}

\subsection{\label{mond}Gravitational interactions in the
deep-MOND regime} I now discuss in more detail matter-TM gravity in
the deep-MOND regime of the symmetric theories. Look at a system of
matter ($\r$) and TM ($\rh$) where the surface densities are
everywhere small enough that the accelerations everywhere are much
smaller than $\az$.\footnote{For our results to apply it is enough
that the accelerations are small across the typical inter-particle
distance in the system. For example the bodies we treat may be point
masses, so the acceleration field near them is higher than $\az$.
All we require then is that the Newtonian regimes of the different
masses stay far away from each other. The point masses can then be
taken as spheres larger then their MOND radii.} In this case, the
potential field is determined from eq.(\ref{katmer}), and the
potential felt by matter is $\f$ itself while that felt by TM is
$-\f$. To the MOND forces determined from these potentials one adds
the subordinate Newtonian forces, coming from the $\ft$ potential,
weighted by $\z$, as per eq.(\ref{rabutret}).
\par
Consider now forces on non-test bodies of both types. In Milgrom
(1997), I showed that the theory described by eq.(\ref{katmer}) is
invariant under conformal transformations, just as the
two-dimensional Poisson equation is. This enables us to derive some
very useful exact results, which are otherwise difficult to obtain
in nonlinear theories of the type discussed here. I also note, in
passing, that the full symmetry of the theory is then the same as
the isometry of a de Sitter space-time, with possible ramifications
discussed in Milgrom (2009d). In addition to the usual invariance to
translations and rotations, which have the usual consequences (e.g.
conservation of momentum and angular momentum), the conformal
symmetry in 3-dimensional Euclidean space includes inversions about
a sphere of any radius $R$, centered at any point $\vr\_0$. Namely,
transformations of the form
$\vr\rar\vr'=\vr\_0+R^2|\vr-\vr\_0|^{-2}(\vr-\vr\_0)$; and also
dilatations $\vr\rar\vr'=\l\vr$. The symmetry means that if
$\f(\vr)$ is the MOND potential for mass distributions
$\r(\vr),~\rh(\vr)$, then $\f\^t(\vr')=\f(\vr)$ is the MOND
potential for mass distributions
$\r\^t(\vr')=J^{-1}\r(\vr),~\rh\^t(\vr')=J^{-1}\rh(\vr)$, where
$J(\vr')$ is the Jacobian of the transformation
$J=||\partial\vr'/\partial\vr||$. For example, from dilatation
invariance follows that if $\f(\vr)$ is the MOND potential for
$\r(\vr),~\rh(\vr)$, then $\f(\vr/\l)$ is the MOND potential for
mass distributions $\l^{-3}\r(\vr/\l),~\l^{-3}\rh(\vr/\l)$.
\par
The existence of gravitational masses with opposite signs fits very
naturally in such a conformally invariant theory. To take advantage
of the symmetry, the point at infinity has to be treated on a par
with all other points, since inversions, which are conformal
transformations, interchange the point at infinity with a finite
point. This implies that we should view the Euclidean space as the
3-dimensional sphere from the topological point of view. On the
other hand, the gravitational field lines going to infinity away
from a system of finite total mass, $M$, converge at infinity with a
net flux, implying an effective negative mass $-M$, there. Under
inversions this mass is brought to a finite point, leaving us with a
configuration involving a negative mass, even if we start without
one. So, having such masses from the start is an advantage. Put
differently: a compact manifold such as the Euclidean sphere, as our
space has to be viewed in a conformally invariant theory (here, to
space transformation, not space-time), cannot accommodate a finite
total mass: Applying Gauss theorem to any surface that separates
space in two, implies that the total mass on one side has to be
equal and opposite that on the other side, hence negative masses are
needed.
\par
Next, I describe several corollaries of the conformal symmetry. They
pertain to a system made of matter and TM point masses, $\mi,~\mhk$
(all defined as positive), at positions $\vr\_i,~\hat\vr\_k$, the
forces on which are $\vF\_i,~\hat\vF\_k$. Translation invariance
dictates $\sum\_i\vF\_i+\sum\_k\hat\vF\_k=0$, and rotational
invariance implies
$\sum\_i\vr\_i\times\vF\_i+\sum\_k\hat\vr\_k\times\hat\vF\_k=0$.
\par
(i) A `virial relation' holds, which takes the form
 \beq \sum\_i\vr\_i\cdot\vF\_i+\sum\_k\hat\vr\_k\cdot\hat\vF\_k=
 -{2\over 3}
 (\az G)\^{1/2}[|\sum\_i\mi-\sum\_k\mhk|\^{3/2}
 -\sum\_i\mi\^{3/2}-\sum\_k\mhk\^{3/2}].   \eeqno{luprat}
There is an additional relation that applies for systems with
vanishing total `charge' (see Milgrom 1997).
\par
(ii) The following are corollaries of relation (\ref{luprat}) for
the two body case: For two matter masses, say $M$ at the origin and
$m$ at $\vr$, the force on $m$ is
 \beq\vF=-{2\over 3}
 (\az G)\^{1/2}[(M+m)\^{3/2}
 -M\^{3/2}-m\^{3/2}]{\vr\over r^2}; \eeqno{mitaq}
so $f(q,\l\gg 1)\approx(2/3)\l^{-1}[(1+q)\^{3/2}-1-q\^{3/2}]$. This
force, which is the same for two TM masse, is always attractive. For
a matter mass $M$ at the origin and a TM mass $\hat M$ at $\vr$, the
(repulsive) force on $\hat M$ is
 \beq \hat\vF={2\over 3}
 (\az G)\^{1/2}[M\^{3/2}+\hat M\^{3/2}-|M-\hat M|\^{3/2}]{\vr\over
r^2};\eeqno{loreb} so, $f\^*(q,\l\gg
1)\approx(2/3)\l^{-1}[1+q\^{3/2}-(1-q)\^{3/2}]$. For $\hat M=M$,
$\hat\vF=(4/3)(\az G)\^{1/2}M\^{3/2}(\vr/r^2)$. This is stronger by
a factor $1/(2\^{1/2}-1)\approx 2.4$ than the attracting force
between two equal matter or TM masses.
\par
(iii) The theory is nonlinear; so forces are not additive: When we
have more bodies present, the force on each has to be determined
through a new calculation of the whole system. This, to my
knowledge, is impossible to derive analytically, in general, but
there are exception, several examples are discussed below to
demonstrate that matter-TM repulsion is general (see also Milgrom
2002a for general theorems on this issue).

\par
(iv) The forces in a three-body system with `vanishing total
charge'; i.e., with $\sum\_i\mi=\sum\_k\mhk$, was calculated in
Milgrom (1997). The force on one of the bodies, call it 1, is
 \beq \vf\_1=\b\_{12}{\vr\_2-\vr\_1\over |\vr\_2-\vr\_1|^2}+
\b\_{13}{\vr\_3-\vr\_1\over |\vr\_3-\vr\_1|^2}, \eeqno{gutes}
 where
$\b\_{ij}=(2/3)(\az
G)^{1/2}(|q\_i+q\_j|\^{3/2}-|q\_i|\^{3/2}-|q\_j|\^{3/2})$, with
$q\_i=\mi$ for a matter body, and $q\_k=-\mhk$ for a TM body. So,
interestingly, in this case the force is simply the sum of the two
forces that would have applied by the two bodies separately.
\par
(v) Other configurations for which there are analytic results
involve symmetric configuration. For example, take a configuration
of $N$ equal masses $M$, of the same type, symmetrically placed at a
distance $r$ from the origin, so that the forces on all are equal
and radial (such as at the corners of a square, a cube, etc., or
forming a uniform spherical shell in the limit of large $N$).
Application of relation (\ref{luprat}) give the force on each mass
as $-(2/3)M(MG\az)^{1/2}(N^{3/2}-N)\vr r\^{-2}$. Now place any mass
$m$ of the opposite type at the center and the force becomes
$-(2/3)(G\az)^{1/2}(|NM-m|^{3/2}-NM^{3/2}-m\^{3/2})\vr r\^{-2}$. The
addition of $m$ clearly weakens the self attraction towards the
origin, and reverses the sign of the force for large enough $m$.
\par
In a similar fashion, if we have an equal number of matter and TM
point bodies of the same mass arranged symmetrically, so that all
forces act radially and are equal (for example two masses of each
type on alternate corners of a square), relation (\ref{luprat})
tells us that they are all repelled from the center.

\par
(vi) The potential field of a pair of equal matter and TM masses,
$M$, at $\vr\_1$ and $\vr\_2$, respectively (in the MOND regime) is
 \beq \f(\vr)=(MG\az)^{1/2}ln{|\vr-\vr\_1|\over|\vr-\vr\_2|}.
 \eeqno{gapusha}
\par
(vii) More generally, all the physics of a system of gravitating
point `charges', $q\_i$, is encapsuled in the $N$-mass energy
functions $\E$, since the forces are derived from them through
$\vF\_i=-\partial E/\partial\vr\_i$ (The force on a given mass
measures the change in the energy under a rigid translation of the
mass.)\footnote{The gravitational energy of an isolated system with
non-vanishing total `charge' is infinite (it is finite for a
vanishing total `charge'), but its difference for two systems with
the same `charge' are finite, and only these interest us.} The above
results give us the analytic expressions for $E$ for the general
two-body system:
 \beq
E(\vr\_1,\vr\_2)=ln|\vr\_1-\vr\_2|\^{\b\_{12}},\eeqno{koiple}
 and for a
three-body system with vanishing total `charge':
 \beq E(\vr\_1,\vr\_2,\vr\_3)=ln[r\^{\b\_{12}}\_{12}
 r\^{\b\_{13}}\_{13}  r\^{\b\_{23}}\_{23}], \eeqno{turgal}
where $r\_{ij}=|\vr\_i-\vr\_j|$  [Relation (\ref{koiple}) is, in
fact, a special case of this with a renormalized energy for the
limit where mass 3 is sent to infinity ($\vr\_3\rar\infty$)]. We do
not know the general form of $E$. For systems with vanishing total
`charge' we can show that $\E$ is determined up to an unknown
function of the $N(N-3)/2$ conformally invariant variables
$u\_{ijkl}=r\_{ij}r\_{kl}/r\_{ik}r\_{jl}$:
 \beq \E=E_0(u\_{ijkl})+\sum\_{1\le i<j\le N}b\^N\_{ij} ln~r\_{ij},
  \eeqno{milesa}
where (for $N>2$)
 \beq b\^N\_{ij}={2\over (N-2)}\left[-\n(q\_i) -\n(q\_j)
 +{1\over(N-1)}\sum\_{m=1}\^N\n(q\_m)
\right], \eeqno{miretab} with $\nu(q)=(2/3)(G\az)^{1/2}|q|^{3/2}$.
For $N=3$ we have no $u\_{ijkl}$ variables, so $E_0$ is a constant
and $\b\_{ij}=b^3\_{ij}$. Both $E$ and $E_0$ have a suppressed
dependence on the charges.\footnote{In $D$ dimensions $\n(q)\propto
q\^{D/(D-1)}$. In two dimensions we have the exact result
$\E\propto\sum\_{i<j} q\_iq\_j ln~r\_{ij}$.} The second term in
eq.(\ref{milesa}) carries the anomalous transformation properties of
the energy under dilatations and special conformal transformations,
while $E_0$ is truly invariant.\footnote{The energy of a
zero-total-charge system is invariant to all conformal
transformations of the charge distribution. However, here we speak
of transforming the point charges rigidly to their new positions,
without affecting the transformation on their internal structure.
The anomalous transformation properties of $\E$ result from this and
the fact that the energy of a finite-charge system does change under
dilatations, $\vr\rar\l\vr$,  by $\n(q)ln\l$.} For example, under
dilatations
  \beq\El{\l}=\E-ln~\l\sum\_{m=1}\^N\nu(q\_m). \eeqno{hupner}
Again, by sending one of the charges to infinity (call it
$q\_N=-Q$), and renormalizing the energy by subtracting the constant
$ln(r\_N)\sum\_1\^{N-1}b\^N\_{iN}=-2\n(Q)ln(r\_N)$, which goes to
infinity in the limit, we can define the energy for a system of
charges with a finite total charge $Q$; for example, a system with
only matter masses (good only for comparing systems with this total
charge). The sum in eq.(\ref{milesa}) then goes only up to $N-1$,
and the $u$ variables with index $N$ are written, e.g.,
$u\_{ijkN}=r\_{ij}/r\_{ik}$; so they are still invariant under
dilatations (but not inversions); so $E_0$ is invariant under
dilatations of $\vr\_1,...,\vr\_{N-1}$. The energy for the remaining
$N-1$ masses transforms under dilatations as:
 \beq E(\l\vr\_1,...,\l\vr\_{N-1})=E(\vr\_1,...,\vr\_{N-1})
 +Aln~\l, \eeqno{hupnam}
 where
  \beq A=\sum\_{1\le i<j\le N-1}b\^N\_{ij}
  =\nu(Q)-\sum\_{m=1}\^{N-1}\nu(q\_m); \eeqno{yertis}
this is equivalent to eq.(\ref{luprat}) (seen by taking the $\l$
derivative at $\l=1$).

\par
(viii) In general, the forces in two systems that are related by a
conformal transformation of the charges are simply related (see
Milgrom 1997). For example, consider a point charge $q$ at a
distance $r<a$ from the center of a spherical shell of radius $a$,
uniformly charged by $-q$. On dimensional grounds, the force on the
point mass can be written as $f(r)=-2\n(q)r a\^{-2}s(r/a)$. Then,
from conformal invariance, the force on $q$ when it is at $R>a$ is
$f(R)=2\n(q)R\^{-1}[1+ (a/R)\^2s(a/R)]$ (the function $s$ is not
known).
\par
Under conformal transformations, equipotential surfaces go to
equipotential surfaces of the new configuration, and field lines go
to field lines. Also, spheres (circles) go to spheres (circles),
including planes (straight lines), which are spheres of infinite
radius. For an arbitrary system of charges, $q\_i$, with $\sum
q\_i=0$, lying on a sphere of radius $a$, one can show\footnote{This
is done by transforming the sphere into a plane where the forces
must all lie from symmetry, and using the transformation law for
forces from Milgrom (1997).} that the radial component of the force
on $q\_i$ is $\n(q\_i)/a$ (pointing outwards), and everywhere on the
sphere, outside the charges, the field lines are tangent to the
sphere (no radial force on test charges).

\subsection{\label{well}Well mixed configurations}
In subsection \ref{mond}, I considered mainly systems of point
masses for which matter and TM are well separated. It is worth
noting that interesting situations may occur when matter and TM are
well mixed in the sense that $\rh\approx \r$.
\par
Look, for example, at the case $\rh=\r$: In this case $\fb=0$, and
$\ft$ is twice the Newtonian potential for each mass separately,
call it $\fN$. So the two sectors see the same potential,
$\f=\fh=2\z\fN$. Given also the same initial velocities, such mass
configurations will retain their density equality with time, with
each mass type developing according to Newtonian dynamics with an
effective Newton constant $G/\b$. This echoes the situation in
symmetric cosmology, where we have $\hgmn=\gmn$, with each
satisfying Freedmann's equations [possibly with a cosmological
constant $\sim\M(0)\az^2$] and with an effective gravitational
constant $G/\b$. Such local configurations are, however, unstable,
and a small departure from equal densities will lead to eventual
segregation of the two mass types. For example, in the presence of a
small matter, or TM, mass $\r\^*$ (or with a small departure from
$\r=\rh$) a nonzero $\fb$ potential is created which is added to
$\f$ and $\fh$ with opposite signs, causing the separation of $\r$
from $\rh$ in a way that increases the separating force even
further. Thus initially, at least, even if the surface densities of
$\r$ and $\rh$ separately are large compared with $\az/G$, so that
each mass type creates a high-acceleration Newtonian field, the
resulting evolution of the difference $\d=\r-\rh$ will be governed
by MOND, and is nonlinear.
\subsection{\label{black}Relativistic point mass solutions}
Allowing for TM in BIMOND permits a two-parameter ($M,~\hat M$)
family of (relativistic) spherically symmetric, static, vacuum
solutions on a double Minkowski background (neglecting `cosmological
constant' effects). These would describe, e.g., matter-TM black
holes. For $M=\hat M$ we have the solutions $\hgmn=\gmn$, with both
metrics being of the Schwarzschild form for $M$, with $G/\b$ as
gravitational constant. At the other end, for $\hat M=0$ we can have
pure-matter black holes, for instance. Because of the cosmological
coincidence $\az\approx cH_0/2\pi$, for all sub-Universe systems the
horizon acceleration is much larger than $\az$. This means that MOND
effects enter only far beyond the horizon, deep in the NR regime.
For $\b=1$ we know that in the high-acceleration regime BIMOND
describes two separate Einstein theories for matter and TM [with a
cosmological constant $\sim\az^2\M(\infty)$, which I neglect here].
In this case, each of the metrics is nearly a Schwarzschild one
corresponding to its own mass, within many Schwarzschild radii. At
large radii the metrics approach their NR expressions as per
eqs.(\ref{rukun})(\ref{katpet}-\ref{rabutret}): at first $\f\approx
-MG/r$, $\fh\approx -\hat MG/r$, and only at much larger radii
$\f\approx-\fh\approx\fb\approx [(M-\hat M)G\az]^{1/2}ln~r$ (for
$M\ge \hat M$).

\subsection{\label{zeromass}Asymptotic field of equal-mass systems}
For an isolated system with equal total masses of matter and TM,
$M=\hat M$, the asymptotic behavior of the MOND potential is not
described by the usual logarithmic dependence on the radius. This
can be easily seen by applying the Gauss theorem to
eq.(\ref{katmer}), which implies a vanishing coefficient to a
logarithmic term. Instead, the generic asymptotic potential is of
the form
 \beq \f(\vr)\rar (MG\az)^{1/2}\vR\cdot{\vr\over r^2},  \eeqno{dipola}
where $\vR$ is some radius vector characteristic of the mass
distribution in the system. This is the vacuum solution of
eq.(\ref{katmer}) with the slowest decrease with radius. It is
gotten by a conformal transformation (inversion at the origin) from
the configuration of a constant acceleration field $\f\propto
\vR\cdot\vr$, which is clearly a vacuum solution, and is thus itself
a vacuum solution by virtue of the conformal invariance of
eq.(\ref{katmer}) (Milgrom 1997).\footnote{The logarithmic vacuum
solution transforms to itself.} Not surprisingly, the potential in
eq.(\ref{dipola}) has the form of a dipolar electrostatic field in
two dimensions: $\vD\cdot\vr/r^2$. However, here the asymptotic
``dipole strength'' is not related to the dipole of the mass
distribution; in fact, I do not know how to express $\vR$ in terms
of the mass distribution. For a matter-TM point-mass pair, the exact
expression, eq.(\ref{gapusha}), implies $\vR=\vr\_2-\vr\_1$. More
generally, we know that  $\vR$ does not scale with the mass, and
scales with the system size: when $\r(\vr)\rar\l^{-3}\r(\vr/\l)$, we
have $\vR\rar\l\vR$.
\par
The magnitude of the asymptotic field in eq.(\ref{dipola}),
$|\gf|=(MG\az)^{1/2}|\vR|r^{-2}$, depends only on $r$ (not on the
direction) and decreases with radius like the Newtonian field of the
system $|\gft|=2MGr^{-2}$. Their ratio is
$|\gft|/|\gf|=2R\_M/|\vR|$, where $R\_M$ is the MOND radius of the
system. So, if the mass distribution is such that $|\vR|\gg R\_M$,
the MOND contribution dominates the field asymptotically as well.
\par
There are configurations for which $|\vR|$ is much smaller than the
characteristic size of the system. In particular, there are systems
for which $\vR=0$, for example due to symmetry. For these, the MOND
potential decreases faster asymptotically, and the Newtonian field
dominates there. I have not identified the next slowest decreasing
vacuum solution of eq.(\ref{katmer}). Because of the conformal
invariance of the problem, finding such solutions is equivalent to
finding the behavior of vacuum solutions near the symmetry point of
a symmetric mass configuration. Take, for instance, a system with
two matter masses, $M$, at opposite corners of a square, centered at
the origin, and two TM masses $\hat M=M$ at the other two corners.
Inversion about a sphere centered at the origin, and containing the
corners, leaves the mass configuration invariant, but interchanges
the origin with infinity. This implies that the behaviors of the
potential near the origin and at infinity are gotten from each other
by a simple inversion. Spherically symmetric configurations with
$M=\hat M$ have a vanishing MOND potential $\f$ outside the mass,
with only $\ft$ contributing.
\section{\label{observational}Observational consequences}
No BIMOND effects are felt in the cosmos, in the picture in which on
cosmological scales $\hgmn=\gmn$, as long as matter is homogeneous
(except perhaps those of a cosmological constant). BIMOND effects
appear when inhomogeneities develop--the beginning of structure
formation--when the presumably uncorrelated perturbations induced in
the two sectors would have caused local departures from equality of
the metrics. The initial stages of the growth of perturbations on
the background of the expanding Universe would have been affected
already by some of the peculiar matter-TM gravitational interaction
discussed here, but these require using the relativistic version of
the theory. Subsequent, nonlinear development of inhomogeneities
occur already in the NR regime, for which we can directly use the NR
theories, and the results of this paper. Matter-TM interactions can
also lead to observable effects in the present day Universe. Here I
discuss very succinctly some possible consequences of the existence
of TM and its BIMOND interaction with matter.

\subsubsection{Structure formation}
Even if the underlying theory itself is symmetric in the two
sectors, it is not clear that the attributes of their cosmic matter
contents are exactly the same. For example, we do not yet know what
engendered the baryon asymmetry in matter. It is possible, then,
that the small initial asymmetry was different in the two sectors,
which could result in different baryon densities today. Also, even
if there occurred similar inflation episodes in the two sectors,
they could have emerged from them with different perturbation
amplitudes. The effects of such possible asymmetries, and others, on
structure formation should be studied.
\par
The presence of TM and its peculiar interactions with matter,
discussed here, would have had major impact on the process of
structure formation. As already mentioned in the introduction, in a
symmetric universe, with $\hgmn=\gmn$ on large scales, small
fluctuations are governed by nonlinear MOND dynamics, even at early
times when the universe at large is characterized by high
accelerations, $cH\gg\az$. What dictates that we are in the MOND
regime is the smallness of the perturbations, leading to small
departures from metric equality between the two sectors: The
argument of the interaction function $\M$, or its derivative $\M'$
appearing in the field equations, is the difference in the
connections of the two metrics in the relativistic regime, or the
difference in the gradients of the matter and TM potentials in the
NR regime, in units of $\az$. The growth of perturbations in this
picture is thus nonlinear from the start, since the MOND potential
is nonlinear in the overdensity. To apply the present results to the
problem would thus require numerical simulations.
\par
Simulations of structure formation with NR MOND dynamics has been
considered in several studies; e.g., by Sanders (2001), Nusser
(2002), Stachniewicz \& Kutschera (2002), Knebe (2005), and
Llinares, Knebe, \& Zhao (2008). It should be easy to extend such
simulation to reckon with TM, by including an initial TM
distribution with its own seed fluctuations, and employing the field
equations (\ref{katpoy})(\ref{rabutla}), or alternatively the more
wieldy eqs.(\ref{katreq})(\ref{rabutla}).
\par
Our results here are not directly applicable to the early stages of
the growth of perturbations, because they assume that the system at
hand is much smaller than the characteristic cosmological curvature
radius, and they ignore the underlying cosmic expansion, relying on
perturbations around Minkowski metrics, not those describing
cosmology. However, these results would still apply at later times,
and might be indicative of what happens qualitatively even at early
times.
\par
One important effect that can be anticipated is the segregation of
matter and TM due to their mutual repulsion. This would presumably
lead to formation of interweaving cosmic webs for matter and TM;
i.e., complementary networks of nodes, filaments, and voids, with
the high-density nodes of one sector residing in the voids of the
other, and vice-versa, with filaments avoiding each other.
\par
Some of the main questions that can be answered by numerical
simulations are: Do matter and TM indeed segregate efficiently and
what does the present-day configuration look like? Do they indeed
form interweaving, mutually avoiding cosmic webs? Is segregation
practically complete, or do we still find galaxy size or larger TM
bodies lingering in the neighborhood of matter structures? Does this
occur often enough for direct gravitational effects of TM on matter
to be a common phenomenon? To what extent voids are more pronounced
(empty) compared with simulations with matter only? What are the
effects on peculiar motions of large-scale structures?
\par
Regarding the question of void structure, in particular, it has been
suggested that various observed aspects of voids pose problems for
the LCDM paradigm (e.g., Peebles 2001, Tully 2007, Tully \& al.
2008, Tikhonov \& Klipin 2009, Tikhonov \& al. 2009, and see  van de
Weygaert \& Platen 2009, and Peebles \& Nusser 2010 for recent
reviews). Even without TM, MOND is expected to produce more
pronounced voids than DM for the following reason: A void acts like
a region of negative mass. If the ambient density is $\r\_a$, then a
void can have at most the effective density $-\r\_a$. In MOND this
affective density can be much higher if the characteristic
accelerations involved are small enough. TM can help even more
because of the added repulsion by TM concentrations, which would
arguably reside in matter voids.
\par
Simulations should also study the effects of various possible
departures from symmetry between the two sectors, such as disparate
baryon densities, different amplitudes of the initial fluctuations,
etc. One should also investigate the dependence of the outcome on
the parameter $\z$, and also consider structure formation in
asymmetric theories with $\a\not =\b$.
\subsubsection{Gravitational lensing}
The best prospects for detecting or constraining TM, if it has
formed structures like those of matter, and if it has segregated
efficiently from matter, seem to be via weak gravitational lensing.
\par
Expressions (\ref{rukun}) for the two NR metrics imply that we can
calculate lensing in the standard way, as is done in GR, using the
MOND potentials instead of the Newtonian potentials.\footnote{This
is not the case for BIMOND theories that hinge on scalar arguments
of the bimetric interaction other than that given in
eq.(\ref{scalala}).}
\par
We are interested here in the lensing effects of an isolated TM
body, $\rh$, on matter photons. These are dictated by the matter
MOND potential $\f$ created by $\rh$. Use, as an example, the field
equations in the form of eq.(\ref{hugtal}), with $\a=\b=1$, and
$\r=0$. We then have
  \beq \D\f=-\fpg\rph,~~~~\D\fh=\fpg(\rh+\rph), \eeqno{hiresa}
where $\rph=-\div(\M'\gfs)$. In other words, we can calculate $\f$
as the Newtonian potential produced by minus the phantom density
produced by $\rh$ alone. This phantom density is the same as we
would calculate for a matter distribution $\r=\rh$. Thus lensing of
matter photons by a TM body are different in two major ways from
lensing by a matter body of the same mass distribution: First,
photons (or any matter test particle) do not `see' the TM itself,
only the fictitious, MOND, phantom matter that it produces. Second,
photons sense this phantom matter with an opposite sign: a TM lens
thus acts as a diverging lens.
\par
The first fact above implies that photons are oblivious to the
existence of the TM in regions of high acceleration: For a spherical
body with radially decreasing acceleration, no force is felt in the
Newtonian regime. For example, a point TM mass $\hat M$, is not felt
by matter roughly within its MOND radius $R\_M=(\hat M
G/\az)^{1/2}$.
\par
Another fact to recall is that the thin-lens approximation does not
apply in MOND (Mortlock and Turner 2001, Milgrom 2002b, Milgrom \&
Sanders 2008) and structure along the line of sight has to be
reckoned with: What enters lensing is still the integrated surface
(column) density of the phantom matter. This, however, does not
depend only on the surface density distribution of the baryonic
matter that produces it; it also depends crucially on the baryon
distribution along the lines of sight. For simplicity's sake, I
consider here only a spherical lens. In this case, the ratio of the
Einstein radius of a lens to its MOND radius is
 \beq {r\_E\over R\_M}=\left(4\az
d\_{ls}d\_{l}\over c^2d\_{s}\right)^{1/2},  \eeqno{mitopl} with
$d\_{ls},~d\_{l},~d\_{s}$ the lens-source, lens, and source
angular-diameter distances, respectively. Considering the well known
proximity $\az\approx cH_0/2\pi$, this ratio is
 \beq {r\_E\over R\_M}\approx
 \left({2d\_{ls}d\_{l}\over \pi d\_{s}D\_H}\right)^{1/2},
  \eeqno{hutgem}
where $D\_H=c/H_0$ is the Hubble distance. Thus, for lenses nearer
than the Hubble distance, the Einstein radius is within the MOND
radius. But no bending occurs within the MOND radius, since $\hat M$
itself is not felt, and there is no PM there. Also, for a spherical
lens, the PM is by definition always at or outside its MOND radius
and so definitely outside its Einstein radius (the surface density
of PM is always subcritical for a spherical body--see below). We
thus do not expect strong lensing effects of TM on matter
photons.\footnote{With a TM lens that is elongated along the line of
sight we can have strong lensing. For example, consider $N$ equal
point TM masses, $\hat M$, along the line of sight, separated from
each other by more then their individual MOND radius $R\_M$. The
projected PM is then $N$ times the individual contribution, so the
Einstein radius of the system is $N^{1/2}$ times that of a single
mass. For large enough $N$, this is larger than $R\_M$, and we can
then get strong lensing effects. Also, with $\b\not =1$ there will
be attractive or repulsive strong lensing effects.}
\par
Since only the phantom matter affects lensing of matter photons by
TM, we should recall some of its important properties. I mentioned
already that it is to be found only roughly beyond the MOND radius.
Another pertinent attribute is that its local surface density (for a
spherical system) is bounded by some universal value:
$\S\lesssim\Sz\approx\az/2\pi G$. This is related to the observation
of Brada \& Milgrom (1999) that the acceleration produced by the
MOND PM can never much exceed $\az$. More directly, I showed in
Milgrom (2009a) that the MOND prediction for the central surface
density of phantom haloes in spheroidal systems is $\lesssim \Sz$.
This fact is also supported by observations of galaxies (Donato \&
al. 2009, Gentile \& al. 2009). The exact value of the maximum as
deduced from MOND depends somewhat on the MOND formulation and on
the choice of extrapolating function. In comparison with the lensing
critical surface density of the lens $\S_c\equiv c^2d\_{s}/\fpg
d\_{ls}d\_{l}$ we have, using again the numerical proximity of
$\az\approx cH_0/2\pi$,
 \beq {\Sz\over \S_c}={d\_{ls}d\_{l}\over\pi d\_{s}D\_H}.
  \eeqno{vuted}
So PM lensing, and as a result TM lensing of matter photons, is
always sub-critical for spherical lenses, and very much so for low
redshift lenses.
\par
A possible, interesting signature of lensing by the PM of a TM body
is the appearance of a ring, or a projected shell of PM (negative
mass in our case) if the TM body is contained within its MOND
radius, as discussed in Milgrom \& Sanders (2008). With lensing by a
matter body (such as a galaxy cluster) the weak ring feature is
generally masked by the dominant surface density of the baryonic
body. But in the case of lensing by TM, the TM baryons do not
contribute to lensing, and thus do not mask the feature if it is
produced.
\par
We should try then to identify the signature of possible TM bodies,
such as TM galaxy clusters, in weak lensing surveys. Several factors
may turn out to make such identification difficult: In the first
place, we can, of course, not rely on guidance, or substantiating
evidence, from matter effects--as can be done in standard weak
lensing--such as x-ray emission, Sunyaev-Zel'dovich effect, or a
direct view of a galaxy concentration. Second, as we saw, no high
surface density effects are expected, since only the rather
law-surface-density fictitious PM acts as a lens. Third, the signal
should evince the effects of a negative mass concentration, which
may require specialized algorithms to discover. Fourth, matter
structures along the line of sight will act to cancel the lensing
effects of TM bodies, or at least distort and confuse them.
\par
Results of large-scale-structure simulations with TM will pinpoint
the expectations for weak lensing by TM, and help design search
strategies.

\subsubsection{\label{other}Other effects}
If matter-TM segregation is not complete, and galaxy-size TM bodies
are still lurking in the neighborhood of matter galaxies, they may
produce distortions that are visible as warps, lopsidedness, or
ellipticity. In particular, in theories with $\b<1$, matter and TM
attract each other in the high acceleration regime; so, there might
be TM trapped in high acceleration regimes of matter territory. One
may speculate in such a case that small amounts of TM trapped in the
(high acceleration) cores of galaxy clusters might be responsible
for the observed mass discrepancy there (see \ref{betasmall} below).
\par
If structure formation simulations confirm that indeed matter and TM
form interleaving mutually avoiding webs, and if there is matter-TM
symmetry in the cosmos, we can estimate the occurrence of TM
galaxies in matter territory, by seeing how often we find matter
galaxies in voids, which are presumably TM territories.
\par
Clearly, there should also be important effects of the existence of
TM on the appearance of the CMB fluctuations.

\section{\label{asy}Asymmetric theories}
In theories with $\a\not =\b$ we have different gravitational
dynamics in the two sectors. In particular, in such theories the
amount of TM in the universe, and its general properties
(distribution, etc.) can be rather different from those of matter.
Such theories are, clearly, worth investigating, even if they
involve treating less amenable configurations. Here I discuss
briefly the NR limit of several examples of such theories.
\subsection{A theory with $\b=1$ and $\a\gg 1$}
In this example we take the extreme limit $\a\rar \infty$ while
$\b=1$. We can then write the field equations for the MOND
potentials, eqs.(\ref{katpoy}-\ref{rabutla}), as
  \beq \D(\z\ft)=\fpg\rh,~~~~~
 \div\{\Mt'[(\gfb/\az)^2]\gfb\}=\fpg(\r-\rh),  \eeqno{katmul}
 \beq
 \f=\z\ft+\fb,~~~~~~~\fh=\z\ft.
   \eeqno{rabmul}
Since now $\z\rar 0$, we have $\Mt'(\infty)=1$. We find then that
$\z\ft=\fh\^N$ is the Newtonian potential of $\rh$, while $\fb$ is a
solution of the nonlinear Poisson equation
$\div[\m(|\gfb|/\az)\gfb]=\fpg(\r-\rh)$, with $\m$ the standard MOND
interpolating function, and $\f=\fb+\fh\^N$, $\fh=\fh\^N$. Dynamics
in the twin sector is thus fully Newtonian and oblivious to the
matter sector. In contradistinction, the gravitational potential in
the matter sector is governed by the sum of the MOND potential
produced via the nonlinear Poisson equation by $\r-\rh$, with the
Newtonian potential of twin matter.

\subsection{A theory with $\a+\b=0$}
As another example, take the case $\a+\b=0$, which leads to the
quasi-linear formulation of MOND (QUMOND) discussed at length in
Milgrom (2009c). The field equations are then
 $$\D\f=\fpg\b^{-1}\r+\b^{-1}\div(\M'\gf^*)=\fpg(\r+\rp),$$
  \beq \D\fh=\fpg\b^{-1}\rh+\b^{-1}\div(\M'\gf^*)=\fpg(\rh+\rph),
  \eeqno{hugtalad}
where, $\fs=\f-\fh$, and
$$\rp\equiv(\fpg)^{-1}\b^{-1}\div(\M'\gf^*)+(\b^{-1}-1)\r,$$
 \beq
 \rph\equiv(\fpg)^{-1}\b^{-1}\div(\M'\gf^*)+(\b^{-1}-1)\rh.
 \eeqno{juparut}
 Taking the difference of these equations gives
 \beq\D\fs=\fpg\b^{-1}(\r-\rh). \eeqno{myreq}
One first solve the Poison eq.(\ref{myreq}) for $\fs$, and then
another Poisson equation for $\f$ or $\fh$. Accelerations of test
particles are given by $\va=-\gf$, $\hat\va=-\gfh$.
\par
We see that, again, the dynamics within each sector separately are
not the same. If we compare the NR potentials for two configurations
$\rh=\r\_R,~\r=0$, and $\r=\r\_R,~\rh=0$, we see that $\fs$ has an
opposite sign for the two configurations, and hence $\rph\not =\rp$.
\par
The conserved momentum is now $\vP=\int(\r\vv-\rh\hat\vv)$. If we
still define the force on a matter subsystem in the volume
$\upsilon$ as $\vF=-\int\_\upsilon\r\gf\drt$, and that on TM as
$\hat\vF=-\int\_{\hat\upsilon}\rh\gfh\drt$, then for a closed system
made of matter and TM it is $\vF-\hat\vF$ that vanishes, not their
sum. If they do not vanish separately, such an isolated system will
self accelerate (still preserving momentum). This seems paradoxical
if applied to an isolated system, but in the context of a universe
filled with matter and TM in equal amount it does not necessarily
lead to unacceptable behavior since no large scale directed
accelerations can occur. It remains to be checked by numerical
simulations whether this can lead to inconsistencies (conceptual or
observational) in the expected behavior of matter.
\par
Specialize further to $\b=1$, which is particularly transparent: In
this case eq.(\ref{myreq}) implies that $\fs$ is the Newtonian
potential of the system with TM contributing as having negative
(active) gravitational mass. In eq.(\ref{juparut}) we have
$\rp=\rph=(\fpg)^{-1}\div(\M'\gf^*)$. To get a Newtonian limit in
the matter sector we have to have $\M'(z)\rar 0$ for $z\rar\infty$
[$z=(\gfs/\az)^2$]. This also gives standard Newtonian dynamics in
the TM sector. It also means that matter and TM do not interact in
the deep Newtonian regime. In this case the total force on matter
and on TM for a closed system vanish separately.
\par
To get MOND dynamics in the matter sector we have to have
$\M'(z)\approx z\^{-1/4}$ for $z\ll 1$. Thus effectively, it can be
said that we can use a Newtonian calculation with each type of mass
`seeing' the density produced by its own type plus the phantom mass
that is common to both sectors. Because $\rh$ enters the source of
$\fs$ with an opposite sign one may say that the phantom mass
produced by TM alone is repulsive to both types of matter, while
that produced by matter is attractive to both. Thus, in the TM
sector, a point mass $\hat M$ produces a Newtonian, attractive force
on a TM test particle within its MOND radius, $R\_M$, but instead of
the MOND enhancement of gravitational attraction in the matter
sector, in the TM sector MOND effects cause this force to weaken the
attraction, and turn it into repulsion at large distances--a
manifestation of the disparate behavior in the two sectors.
\subsection{\label{betasmall}A theory with
$0<\b\ll 1$ and $\a \ge 1$}
 Here I consider an example with $0<\b<1$, and
to accentuate matters,  with $\b\ll 1$. In such theories matter and
TM attract in the high acceleration limit but still repel in the
deep-MOND limit. Since we have to have $\z<1$ we I take $\a\ge 1$,
or, for concreteness' sake take $\a=1$.
\par
From eq.(\ref{rabutla}) we can now write the potentials for the two
sectors as
 \beq
 \f=\fN+\b^{-1}\hat\f\^N+\fb,
 ~~~~~~~\fh=\fN+\b^{-1}\hat\f\^N-\l\fb,
  \eeqno{rabutlop}
where I used the fact that $\l=\b/\a\ll 1$, and $\fb$ is a solution
of the second of eq.(\ref{katpoy}), i.e.,
$\div\{\Mt'[(\gfb/\az)^2]\gfb\}=\fpg(\r-\rh)$. The gravitational
potential that matter sees is thus the sum of three terms: its own
Newtonian potential, the Newtonian potential of the TM enhanced by
the large factor $\b^{-1}$, and the ``MOND'' potential $\fb$. Note
that because here $\z\approx 1$, we have $\Mt'(\infty)\approx
\l^{-1}\gg 1$, so for $\az\rar 0$, $\fb\rar \l(\fN-\hat\f\^N)$.
\par
We see that even a small admixture of TM in an otherwise pure matter
object can make an important contribution to the matter potential,
because its Newtonian potential is enhanced by the factor $\b^{-1}$.
For example, in such a theory, a small amount of TM trapped in the
cores of (matter) galaxy clusters could explain away the mass
discrepancies observed there. Such amounts of TM need not affect
much the behavior in the MOND regime of the clusters, since TM
enters the source of the second of eq.(\ref{katpoy}) with the same
weight as matter.
\par
It remains to be seen whether such asymmetric theories can be made
consistent with cosmology. Recall that it is $G/\b$ that plays the
role of the Newton constant in the BIMOND cosmologies I have
considered so far. This can be interpreted as an apparent
enhancement of the matter density by a factor $1/\b$. For example,
in the Friedmann equations a baryon density $\r\_b$ appears as
$\r\_b/\b$, which would be interpreted as baryons plus DM of density
$\r\_b(\b^{-1}-1)$.

\section{\label{disc}Summary and discussion}
I have studied some aspects of the dynamics of matter an the
putative TM that may be present in the context of BIMOND. This is
done for NR bodies of either type. In fully symmetric theories,
which seem preferable on various grounds, we get MOND dynamics
within each sector. The interaction between matter and TM is,
however, nonstandard even compared with MOND: In the deep-MOND
regime matter and TM bodies repel each other with MOND-like forces
(decreasing as inverse distance). In the Newtonian,
high-acceleration regime the force depends on the parameter $\b$.
For the fiducial value $\b=1$, the matter-TM interaction vanishes.
For $\b<1$ there is Newtonian-like attraction (decreasing as inverse
squared distance) with strength $\b^{-1}-1$, and for $\b>1$ there is
similar repulsion.
\par
I have also considered briefly possible effects of the presence of
TM with such properties on structure formation and its lensing
properties. To assess more reliably such effects we have to call
upon numerical simulations.
\par
It needs to be stressed, finally, that the TM does not, indeed
cannot, play the full role of dark matter in galactic systems
(although, as we saw, it may produce some effects attributed to dark
matter): There isn't enough of it; it probably shies matter galactic
systems; and, in all probability, would, anyhow, decrease
gravitational attraction between matter bodies if it comes between
them. It is still the MOND departure from standard gravity that is
responsible for the observed mass discrepancy.

\section*{Acknowledgements}
I am grateful to Avi Loeb and Bob Sanders for useful suggestions.
This research was supported by a center of excellence grant from the
Israel Science Foundation.

\clearpage

\end{document}